
\magnification=\magstep1
\baselineskip=9pt

\font\fontC=cmr10

\font\germ=eufm10
\def\ge{\hbox{\germ g}}
\def\ssl{\hbox{\germ sl}}

\def\ttt{\hbox{\germ t}}
\def\llra{\relbar\joinrel\relbar\joinrel\relbar\joinrel\rightarrow}
\def\lan{\langle}
\def\ran{\rangle}

\def\del{\delta}
\def\Del{\Delta}
\def\Delr{\Delta^{(r)}}
\def\Dell{\Delta^{(l)}}
\def\Delb{\Delta^{(b)}}
\def\Deli{\Delta^{(i)}}
\def\ot{\otimes}
\def\nd{\noindent}

\def\lm{\lambda}
\def\Lm{\Lambda}
\def\hom{{\hbox{Hom}}}
\def\vep{\varepsilon}
\def\vp{\varphi}
\def\vpi{\varphi^{-1}}
\def\ovl{\overline}

\def\wtil{\widetilde}
\def\qq{\qquad}
\def\q{\quad}
\def\qi{q_i}
\def\qii{q_i^{-1}}
\def\ti{t_i}
\def\tii{t_i^{-1}}
\def\lar{\rightarrow}
\def\al{\alpha}
\def\uup{U^{\geq}}
\def\ulow{U^{\leq}}
\def\bup{B^{\geq}}
\def\blow{\ovl B^{\leq}}
\def\uq{U_q(\ge)}
\def\catob{{\cal O}(B)}
\def\catobr{{\cal O}^r(B)}
\def\BQ{{\cal B}_q}
\def\UR{{\cal R}}

\def\QQ{\hbox{\bf Q}}
\def\ZZ{\hbox{\bf Z}}
\def\FF{\hbox{\bf F}}
\def\SFF{\scriptstyle\hbox{\bf F}}

\centerline{\bf  QUANTUM R-MATRIX AND}
\centerline{\bf INTERTWINERS FOR THE KASHIWARA ALGEBRA}

\vskip 3mm
\centerline{TOSHIKI NAKASHIMA}
\vskip 3mm
\centerline{Department of Mathematical Science,}
\centerline{Faculty of Engineering Science,}
\centerline{Osaka University, Toyonaka, Osaka 560, Japan }
\centerline{e-mail: toshiki${\fontC\char'100}$sigmath.osaka-u.ac.jp}
\vskip3mm
\centerline{ABSTRACT}

\nd We study the algebra $B_q(\ge)$ presented by Kashiwara
and introduce intertwiners
similar to  $q$-vertex operators. We show that
a matrix determined by 2-point functions of the intertwiners
coincides with a quantum R-matrix (up to a diagonal matrix)
and give the commutation relations of the intertwiners.
We also introduce an analogue of the universal R-matrix for the
Kashiwara algebra.

\vskip 2mm

\beginsection\S0. Introduction

In a recent work [FR], Frenkel and Reshetikhin developed the theory of
$q$-vertex operators. They showed that n-point correlation functions
associated to $q$-vertex operators satisfy
a $q$-difference equation called $q$-deformed Kniznik-Zamolodchikov equation.
In the derivation of this equation, a crucial point is that the quantum
affine algebra is a quasi-triangular Hopf algebra.
By using several properties of the quasi-triangular Hopf algebra and the
representation theory of the
quantum affine algebra, the equation is described in
terms of quantum R-matrices.([FR],[IIJMNT]).

In [K1], Kashiwara introduced the algebra $B^{\vee}_q(\ge)$,
which is generated by 2$\times$rank $\ge$ symbols with the Serre relations and
the $q$-deformed bosonic relations (See Sect.1,(1.5))
in order to study the crystal base of $U^-$, where $U^-$ is a
maximal nilpotent
subalgebra of the quantum algebra $U_q(\ge)$ associated to
a symmetrizable Kac-Moody Lie algebra $\ge$.
(In [K1], $B^{\vee}_q(\ge)$ is denoted by $\BQ(\ge)$).
We shall call this algebra the {\it Kashiwara algebra}.
He showed that $U^-$ has a
$B^{\vee}_q(\ge)$-module structure and it is irreducible. He also showed that
$B^{\vee}_q(\ge)$ has a similar structure to the Hopf algebra:
there is an algebra homomorphism
$B^{\vee}_q(\ge)\lar U_q(\ge)\ot B^{\vee}_q(\ge)$.
Thus if $M$ is a $U_q(\ge)$-module and $N$ is a $B^{\vee}_q(\ge)$-module,
then $M\ot N$ has a $B^{\vee}_q(\ge)$-module structure via this homomorphism.

In the present paper,  we shall first introduce the algebras $B_q(\ge)$,
$\ovl B_q(\ge)$, $U_q(\ge)$ associated to a symmetrizable
Kac-Moody Lie algebra $\ge$ and algebra morphisms for such algebras.
The algebra $B_q$ is obtained
by adding the Cartan part to $B^{\vee}_q$ and
the algebra $\ovl B_q$
is an algebra anti-isomorphic to $B_q$, where we  also call these the
Kashiwara algebras.
The algebra $U_q$ is an ordinary quantum algebra.
The Kashiwara algebra has no Hopf algebra structure,
but these algebras admit  a certain algebra structure similar
 to  the Hopf algebra. In fact,
there are the following algebra homomorphisms,
$U_q\lar U_q\ot U_q$, $B_q\lar B_q\ot U_q$,
$\ovl B_q\lar U_q\ot \ovl B_q$,
$U_q\lar \ovl B_q\ot B_q$,
an antipode $S:U_q\lar U_q$
and an anti-isomorphism $\vp:\ovl B_q\lar B_q$. By using these,
we can consider tensor products and dual modules of
$B_q$-modules, $\ovl B_q$-modules and $U_q$-modules.
(See Sect.1 and Sect.2.)

In Sect.2, we discuss properties of the category of highest weight
$B_q$-modules.
In Sect.3, we recall the Killing form
of $U_q$ due to [R],[T] and give a
certain relationship between the algebra $B^{\vee}_q$
and the Killing form.
We also introduce a bilinear pairing $\lan\,\,|\,\,\ran$
for highest weight $B_q$-module $H(\lm)$, which is
an analogue of an ordinary vacuum expectation value.
In Sect.4, we restrict ourselves
to an affine case and
consider the following type of
intertwiners similar to $q$-vertex operators;
$$
\hom_{B_q}(H(\lm),H(\mu)\widehat\ot V_z), \eqno(0)
$$
and examine the condition for existence of such intertwiners.
By using the bilinear pairing above
for a composition of these intertwiners,
we define 2-point functions.
By using the relationship between the algebra $B^{\vee}_q$ and the
Killing form, we can explicitly describe a 2-point function
as a matrix element of an image of the universal R-matrix.
In other words, 2-point functions give matrix elements of  quantum R-matrix
up to scalar factors.
Here note that we do not derive any type of equation.
This point differs from [FR].
Nevertheless, by pure algebraic method we can describe 2-point functions.

In order to explain precisely, we prepare some notations. Let $U'_q$
be a subalgebra of a quantum affine algebra $U_q$
without a scaling element, let $V$ and $W$ be
finite dimensional $U'_q$-modules, let $V_{z_1}$ and $W_{z_2}$ be their
affinizations, where $z_1$ and $z_2$ are formal variables,
let  $R^{VW}(z_1/z_2)$ be
the image of the universal R-matrix onto $V_{z_1}\ot W_{z_2}$ and let
$u_{\lm}$ (resp. $u^r_{\lm}$) be a highest weight vector of an irreducible
highest weight left (resp. right) $B_q$-module
$H(\lm)$ (resp. $H^r(\lm)$).
\proclaim Theorem.   ( Theorem 5.3.)
For $\Phi_{\lm}^{\mu\, V}(z_1)\in
\hom_{B_q}(H(\lm),H(\mu)\widehat\ot V_{z_1})$
and
$\Phi_{\mu}^{\nu\, W}(z_2)\in
\hom_{B_q}(H(\mu),H(\nu)\widehat\ot W_{z_2})$,
we have
$$
\lan u^r_{\nu}|\Phi_{\mu}^{\nu\, W}(z_2)
\Phi_{\lm}^{\mu\, V}(z_1)|u_{\lm}\ran=q^{(\lm-\mu,\mu-\nu)}
\sigma\,R^{VW}(z_1/z_2)(v_0\ot w_0),
$$
where $\sigma:a\ot b\lar b\ot a$, and $v_0\in V$ and $w_0\in W$
are the leading temrs of
$\Phi_{\lm}^{\mu\, V}(z_1)$ and $\Phi_{\mu}^{\nu\, W}(z_2)$ respectively.
(See Definition 4.1.)
\par
\nd From this theorem and the unitarity of
a quantum R-matrix, we can derive the
commutation relation of intertwiners of type (0).

In section 6, for the algebra
$B_q$ we give an element $\wtil\UR$, which is an
analogue of the  universal R-matrix $\UR$. This satisfies, for example,
$\wtil\UR_{12}\wtil\UR_{13}\UR_{23}=\UR_{23}\wtil\UR_{13}\wtil\UR_{12}$,
{\it etc.}
We also introduce an projector $\Gamma$ associated to $\wtil\UR$,
which acts on $H(\lm)$ and
singles out only the highest weight component.
In Appendix A, we list some formulae for algebra homomorphisms related to the
algebras introduced in this paper
and in Appendix B, we recall the theory of
 the universal R-matrix of $U_q$.

The author would like to acknowledge E.Date, M.Jimbo,  M. Kashiwara and M.Okado
for discussions and helpful advice.

\beginsection\S1. Preliminary

We shall define the algebras playing a significant role in this paper.
First, let $\ge$ be a  symmetrizable Kac-Moody algebra over {\bf Q}
with a Cartan subalgebra $\ttt$, $\{\al_i\in\ttt^*\}_{i\in I}$
 the set of simple roots and
$\{h_i\in\ttt\}_{i\in I}$  the set of coroots,
where $I$ is a finite index set. We define an inner product on
$\ttt^*$ such that $(\al_i,\al_i)\in{\bf Z}_{\geq 0}$ and
$\lan h_i,\lm\ran=2(\al_i,\lm)/(\al_i,\al_i)$ for $\lm\in\ttt^*$.
Set $Q=\oplus_i\ZZ\al_i$, $Q_+\oplus_i\ZZ_{\geq0}\al_i$ and
$Q_-=-Q_+$. We call $Q$ a root lattice.
Let  $P$  a lattice of $\ttt^*$ {\it i.e.} a free
{\bf Z}-submodule of $\ttt^*$ such that
$\ttt^*\cong {\hbox{\bf Q}}\ot_{\ZZ}P$,
and $P^*=\{h\in \ttt|\lan h,P\ran\subset\ZZ\}$.
Now, we introduce the symbols
$\{e_i,e''_i,f_i,f'_i\,(i\in I),q^h\,(h\in P^*)\}$.
These symbols satisfy the following relations;
$$
\eqalignno{
&q^0=1, \q{\hbox{\rm and }}\q q^hq^{h'}=q^{h+h'},&(1.1)\cr
&q^he_iq^{-h}=q^{\lan h,\al_i\ran}e_i,&(1.2a)\cr
&q^he''_iq^{-h}=q^{\lan h,\al_i\ran}e''_i,&(1.2b)\cr
&q^hf_iq^{-h}=q^{-\lan h,\al_i\ran}f_i,&(1.3a)\cr
&q^hf'_iq^{-h}=q^{-\lan h,\al_i\ran}f'_i,&(1.3b)\cr
&[e_i,f_j]=\del_{i,j}(t_i-t^{-1}_i)/(q_i-q^{-1}_i),&(1.4)\cr
&e''_if_j=q_i^{{\lan h_i,\al_j\ran}}f_je''_i+\del_{i,j},&(1.5)\cr
&f'_ie_j=q_i^{{\lan h_i,\al_j\ran}}e_jf'_i+\del_{i,j},&(1.6)\cr
&\sum_{k=1}^{1-{\lan h_i,\al_j\ran}}
(-1)^kX_i^{(k)}X_jX_i^{(1-{\lan h_i,\al_j\ran}-k)}=0,
\q({\hbox{ for }}X_i=e_i,\,e''_i,\,f_i,\,f'_i{\hbox{ and }}i\ne j),&(1.7)\cr}
$$
where $q$ is transcendental over $\QQ$ and
we set $q_i=q^{(\al_i,\al_i)/2}$, $t_i=q_i^{h_i}$,
$[n]_i=(q^n_i-q^{-n}_i)/(q_i-q_i^{-1})$, $[n]_i!=\prod_{k=1}^n[k]_i$
and $X_i^{(n)}=X_i^n/[n]_i!$.

Now, we define the
algebras $B_q(\ge)$, $\ovl B_q(\ge)$ and $U_q(\ge)$. In the rest of this paper,
we denote the base field {{\bf Q}$(q)$}
by $\FF$. The algebra $B_q(\ge)$ (resp. $\ovl B_q(\ge)$)
is an associative algebra generated by the symbols $\{e''_i,f_i\}_{i\in I}$
(resp. $\{e_i,f'_i\}_{i\in I}$) and $q^h$ ($h\in P^*$) with
the defining relations (1.1), (1.2b), (1.3a), (1.5) and (1.7)
(resp. (1.1), (1.2a), (1.3b), (1.6) and (1.7)) over $\FF$.
The algebra $U_q(\ge)$ is an associative algebra
generated by the symbols $\{e_i,f_i\}_{i\in I}$ and $q^h$ ($h\in P^*$)
with the defining relations (1.1),(1.2a),(1.3a),(1.4) and (1.7) over $\FF$.
We shall call algebras $B_q(\ge)$ and $\ovl B_q(\ge)$ the
{\it Kashiwara algebras.}
([K1]).
Furthermore, we define subalgebras
$$
\eqalignno{
&T=\lan q^h|h\in P^*\ran=B_q(\ge)\cap\ovl B_q(\ge)\cap U_q(\ge),\cr
&B^{\vee}_q(\ge)\,({\hbox{resp. }}\ovl B^{\vee}_q(\ge))=
\lan e''_i,\,f_i\,({\hbox{resp. }}e_i,\,f'_i)|i\in I\ran
\subset B_q(\ge)\,({\hbox{resp. }}\ovl B_q(\ge)),\cr
&U^+_q(\ge)\,({\hbox{resp. }}U^-_q(\ge))=
\lan e_i\,({\hbox{resp. }}f_i)|i\in I\ran=\ovl B^{\vee}_q(\ge)\cap U_q(\ge)\,
({\hbox{resp. }}B^{\vee}_q(\ge)\cap U_q(\ge)),\cr
&\uup_q(\ge)\,({\hbox{resp. }}\ulow_q(\ge))=
\lan e_i\,({\hbox{resp. }}f_i),q^h|i\in I,\,h\in P^*\ran
=\ovl B_q(\ge)\cap U_q(\ge)\,({\hbox{resp. }}B_q(\ge)\cap U_q(\ge)),\cr
&B^+_q(\ge)\,({\hbox{resp. }}\ovl B^{\,\,-}_q(\ge))=
\lan e''_i\,({\hbox{resp. }}f'_i)|i\in I\ran\subset B^{\vee}_q(\ge)
\,({\hbox{resp. }}\ovl B^{\vee}_q(\ge)),\cr
&\bup_q(\ge)\,({\hbox{resp. }}\blow_q(\ge))=
\lan e''_i\,({\hbox{resp. }}f'_i),q^h|i\in I,\,h\in P^*\ran
\subset B_q(\ge)\,({\hbox{resp. }}\ovl B_q(\ge)).\cr}
$$
We shall use the abbreviated notations
$U$, $B$, $\ovl B$, $B^{\vee}$,$\cdots$
for $\uq$, $B_q(\ge)$, $\ovl B_q(\ge)$, $B^{\vee}_q(\ge)$,$\cdots$
if there is no confusion.

For $\beta=\sum m_i\al_i\in Q_+$ we set $|\beta|=\sum m_i$ and
$$
U^{\pm}_{\pm\beta}=\{u\in U^{\pm}|q^huq^{-h}
=q^{\pm\lan h,\beta\ran}u\,\,(h\in P^*)\},
$$
and call $|\beta|$ a height of $\beta$ and
$U^+_{\beta}$ (resp. $U^-_{-\beta}$) a weight space of
$U^+$ (resp. $U^-$) with a weight $\beta$ (resp. $-\beta$). We also define
$B^+_{\beta}$ and $\ovl B^{\,\,-}_{-\beta}$ by the similar manner.

We shall define weight completions of $L^{(1)}\ot\cdots\ot L^{(m)}$,
where $L^{(i)}=B$ or $U$.(See[T])
$$
\eqalign{
\widehat L^{(1)}\widehat\ot\cdots\widehat\ot\widehat L^{(m)}=
\lim_{\longleftarrow\atop l}L^{(1)}\ot\cdots\ot L^{(m)}
  /(L^{(1)}\ot\cdots\ot L^{(m)})L^{+,l},\cr}
$$
where
$L^{+,l}=\oplus_{|\beta_1|+\cdots+|\beta_m|\geq l}
{L^{(1)}}^+_{\beta_1}\ot\cdots\ot {L^{(m)}}^+_{\beta_m}$.  (Note that
 $U\cong U^-\ot T\ot U^+$ and $B\cong U^-\ot T\ot B^+$. )
The linear maps $\Del$, $\Delr$, $S$, $\vp$, multiplication, {\it e.t.c.} are
naturally extend for such completions.

\vskip3mm
\nd{\sl Remark 1.1.}
The algebra $B^{\vee}$  is  introduced in [K1]
for studying the  crystal base of $U^-$ and called the reduced $q$-analogue.
Note that in [K1] the algebra defined by
the relation
$e'_if_j=q_i^{-{\lan h_i,\al_j\ran}}f_je'_i+\del_{ij}$ is mainly studied, but
there is no essential difference since both are equivalently related to each
other by $q\leftrightarrow q^{-1}$.

We shall introduce the algebra homomorphisms related to the algebras
defined above.
\proclaim Proposition 1.2.
(1) If we define linear maps $\Del:U\lar U\ot U$,
$\Delr: B\lar B\ot U$, $\Dell:\ovl B\lar U\ot\ovl B$
and $\Delb:U\lar\ovl B\ot B$ by
$$
\eqalignno{
&\Del(q^h)=\Delr(q^h)=\Dell(q^h)=\Delb(q^h)=q^h\ot q^h,&(1.8)\cr
&\Del(e_i)=e_i\ot 1+\ti\ot e_i,\qq
\Del(f_i)=f_i\ot\tii+1\ot f_i,&(1.9)\cr
&\Delr(e''_i)=(\qi-\qii)\ot\tii e_i+e''_i\ot\tii,\qq
\Delr(f_i)=f_i\ot\tii+1\ot f_i,&(1.10)\cr
&\Dell(e_i)=e_i\ot 1+\ti\ot e_i,\qq
\Dell(f'_i)=(\qi-\qii)\ti f_i\ot 1+\ti\ot f'_i,&(1.11)\cr
&\Delb(e_i)=\ti\ot{{\ti e''_i}\over{\qi-\qii}}+e_i\ot 1,\qq
\Delb(f_i)=1\ot f_i+{{\tii f'_i}\over{\qi-\qii}}\ot\tii,&(1.12)\cr}
$$
and extending these to the whole algebras
by the rule: $\Del(xy)=\Del(x)\Del(y)$
and $\Deli(xy)=\Deli(x)\Deli(y)$ ($i=r,l,b$),
then they give  well-defined algebra homomorphisms.\hfill\break
(2) If we define linear maps $S:U\lar U$ and
$\vp:\ovl B\lar B$ by
$$
\eqalignno{
&S(e_i)=-\tii e_i,\qq S(f_i)=-f_i\ti,\qq S(q^h)=q^{-h}&(1.13)\cr
&\vp(e_i)=-{1\over{\qi-\qii}}e''_i,\qq
\vp(f'_i)=-(\qi-\qii)f_i,\qq \vp(q^h)=q^{-h},&(1.14)\cr}
$$
and extending these to the whole algebras
by the rule: $S(xy)=S(y)S(x)$ and $\vp(xy)=\vp(y)\vp(x)$,
then these maps give well-defined anti-isomorphisms.

\nd Note that in [K1] a homomorphism similar to $\Delr$ is introduced.

\nd{\sl Proof.}
By direct calculations, we can check all the commutation relations. But
it is too complicated to check the Serre relations directly. Since the map
$\Del$ is an ordinary comultiplication,
we may assume that $\Del$ is well-defined.
The formulae (A10), (A11), (A12) and (A13) in Appendix A are
useful for checking the Serre relations. For example, from (A10)
and  the fact: $\Dell_{|\uup}=\Del_{|\uup}$, we have
$$
\eqalignno{
&\Delr(\sum_{k=1}^{1-{\lan h_i,\al_j\ran}}(-1)^k{e''_i}^{(k)}{e''_j}
{e''_i}^{(1-{\lan h_i,\al_j\ran}-k)})
=\sigma(S\ot \vp)\Dell\vpi
(\sum_{k=1}^{1-{\lan h_i,\al_j\ran}}
(-1)^k{e''_i}^{(k)}{e''_j}{e''_i}^{(1-{\lan h_i,\al_j\ran}-k)})\cr
&=(q^{-1}_j-q_j)(\qii-\qi)^{1-{\lan h_i,\al_j\ran}}\sigma(S\ot \vp)\Del
(\sum_{k=1}^{1-{\lan h_i,\al_j\ran}}
(-1)^ke_i^{(k)}e_je_i^{(1-{\lan h_i,\al_j\ran}-k)})=0.
&{\hbox{Q.E.D.}}\cr}
$$

\vskip3mm
\nd{\sl Remark 1.3.}
\item{(1)} If we define an algebra homomorphism $\vep:U\lar \FF$ by
 $\vep(e_i)=\vep(f_i)=0$ and $\vep(q^h)=1$,
then $(\Del,S,\vep)$ gives a Hopf algebra structure on $U$.
\item{(2)} The following diagrams are commutative:
$$
\matrix{
&B&\buildrel{\Delr}\over\llra& B\ot U\cr
&{{\scriptstyle\Delr}\downarrow\,\,\,\,}
&&{\scriptstyle 1\ot\Del}\downarrow\qq\cr
&B\ot U&\buildrel{\Delr\ot1}\over\llra& B\ot U\ot U\cr}\qq
\matrix{
&\ovl B&\buildrel{\Dell}\over\llra& U\ot\ovl B\cr
&{{\scriptstyle\Dell}\downarrow\,\,\,\,}
&&{\scriptstyle \Del\ot1}\downarrow\qq\cr
&U\ot\ovl B&\buildrel{1\ot\Dell}\over\llra& U\ot U\ot\ovl B\cr}
$$
Thus for a $B$ (resp. $\ovl B$)-module $L$, and $U$-modules $M$ and $N$,
there is an isomorphism
of $B$ (resp. $\ovl B$)-module;
$$
(L\ot M)\ot N\cong L\ot (M\ot N)\q
({\hbox{resp.}} (M\ot N)\ot L\cong M\ot(N\ot L)).
$$
Hence we write these $L\ot M\ot N$ (resp. $M\ot N\ot L$). More generaly,
if  $M$ is a $B$ (resp. $\ovl B$)-module
and $N_1,\cdots,N_k$ are $U$-modules,
then $M\ot N_1\ot\cdots\ot N_k$ (resp. $N_1\ot\cdots\ot N_k\ot M$) is a
well-defined $B$ (resp. $\ovl B$)-module.
\item{(3)} If $M$ is a $\ovl B$-module and $N$ is a $B$-module,
then $M\ot N$
has a $U$-module structure via $\Delb$.
\item{(4)} From (A8) (resp. (A9) ) and the coassociative laws
of $\Delr$ (resp. $\Dell$) and $\Del$ as in (2), we know
that $B$ (resp. $\ovl B$) has a
right(resp. left) $U$-comodule structure. (see [A].)
\item{(5)} The algebra $\bup$ (resp. $\blow$) is isomorphic to
$\uup$ (resp. $\ulow$) as an associative algebra, but $\bup$ (resp. $\blow$)
has no Hopf algebra structure, thus we do not identify them.

\vskip2mm
\nd We list several formulae for these operations in Appendix A.

\beginsection\S2 Representation theory of the Kashiwara algebra

We shall discuss the representation theory of the algebra $B_q(\ge)$.
In the rest of this paper, we
 assume that all representations below have a weight space
decomposition and each weight space is  finite dimensional,
where for a vector space $M$ with a $T$-module structure, a weight space
$M_{\lm}$ with weight $\lm\in \ttt^*$ is defined by
$\{u\in M|q^hu=q^{\lan h,\lm\ran}u\,\,(h\in P)\}$.

\vskip2mm
\nd{\sl 2.1. Dual modules\q}
Let $M$ be a left $B$-module and $h:\ovl B\lar B$  an anti-isomorphism
({\it e.g.} $\vp$  in Sect.1).
Then the dual space $M^*={\hbox{Hom}}_{\SFF}(M,\FF)$
has a left $\ovl B$-module structure by
$$
(xu,v)=(u,h(x)v),\q {\hbox{for }}x\in \ovl B,\,u\in M^*,\,v\in M.\eqno(2.1)
$$
We denote it by $M^{*h}$. Similarly, for a
$\ovl B$-module $N$ and an anti-isomorphism $g:B\lar \ovl B$,
the dual space $N^*$ has a left $B$-module structure and we denote it
by $N^{*g}$.

Let $M$ be a $\ovl B$-module, $N$ be a $U$-module and $g$ be as above.
Then we can give a left $B$-module structure on
${\hbox{Hom}}_{\SFF}(M,N)$ by
$$
(xf)(u)=\sum x_{(2)}f(g(x_{(1)})u),\q{\hbox{for }}x\in B,\,f\in
{\hbox{Hom}}_{\FF}(M,N),\,u\in M,\eqno(2.2)
$$
where we denote $\Delr(x)=\sum x_{(1)}\ot x_{(2)}\in B\ot U$.

\nd Note that there is an isomorphism as a $B$-module;
$$
{\hbox{Hom}}_{\FF}(M,N)\cong M^{*g}\ot N.\eqno(2.3)
$$

Similarly, for  $B$-modules  $M$ and $N$,  we
give a $U$-module structure on $\hom_{\FF}(M,N)$ by;
$$
(yf)(u)=\sum y_{(2)}f(h(y_{(1)})u),\q{\hbox{for }}y\in U,\,f\in
{\hbox{Hom}}_{\FF}(M,N),\,u\in M,
$$
where $\Delb(y)=\sum y_{(1)}\ot y_{(2)}\in \ovl B\ot B$.

\proclaim Proposition 2.1.
Let $L$ be a $\ovl B$-module,  $M$ be a $B$-module,  $N$ be
a $U$-module and $\vp:\ovl B\lar B$ be as in Sect.1.
Then we obtain an isomorphism of vector spaces;
$$
{\hbox{Hom}}_U(L\ot M,N)\cong {\hbox{Hom}}_B(M,\hom_{\FF}(L,N)).\eqno(2.4)
$$

\nd Remark that $L\ot M$ has a $U$-module structure via $\Delb$ and
$\hom_{\FF}(L,N)$ has a $B$-module structure via $\Delr$
according to (2.2).

\vskip2mm
\nd{\sl Proof.} We define a map $\Phi:
{\hbox{Hom}}_U(L\ot M,N)\lar{\hbox{Hom}}_B(M,\hom_{\FF}(L,N))$
as follows: for $f\in {\hbox{Hom}}_U(L\ot M,N)$,
$\Phi(f)$ is given by
$$
\matrix{
\Phi(f)(y):&L\longrightarrow & N\cr
     &x\longmapsto&f(x\ot y)\cr},\q{\hbox{ for }}y\in M.
$$
First we check the well-definedness of
$\Phi$ {\it i.e.} $B$-linearity of $\Phi(f)$.
For $P\in B$, $x\in L$ and $y\in M$ by the definition of $\Phi$, we get
$(\Phi(f)(Py))(x)=f(x\ot Py)$. From (2.2) we can act $P$ on $\Phi(f)(y)$
as follows;
$$
\eqalign{
(P\Phi(f)(y))(x)&=\sum P_{(2)}\Phi(f)(y)(\vp^{-1}P_{(1)}x)\cr
          &=\sum P_{(2)}f(\vp^{-1}P_{(1)}x\ot y)\cr
          &=\sum f(P_{(2)}\vp^{-1}P_{(1)}x\ot P_{(3)}y)\cr
          &=\sum f(\vp^{-1}(P_{(1)}\vp P_{(2)})x\ot P_{(3)}y),\cr}\eqno(2.5)
$$
where $(1\ot\Delb)\Delr(P)=\sum P_{(1)}\ot P_{(2)}\ot P_{(3)}$.
{}From (A2) in Appendix A, the last formula in (2.5) is equal to $f(x\ot Py)$.
Hence $\Phi(f)$ is $B$-linear. The injectivity of $\Phi$ is trivial.
For $k\in \hom_{B}(M,\hom_{\FF}(L,N))$,
we define $\Psi(k)\in \hom_{U}(L\ot M,N)$ by $\Psi(k)(x\ot y)=(k(y))(x)$
($x\in L$, $y\in M$). We can easily check the well-definedness of $\Psi$
and $\Phi\circ\Psi(k)=k$.  \hfill{Q.E.D.}

\vskip 2mm
{}From Proposition 2.1 and (2.3), for a $B$-modules $L$, $M$
and a $U$-module $N$, there is an isomorphism;
$$
\hom_{U}({}^rL^{*\vp}\ot M,N)\cong\hom_{B}(M,\widehat L\ot N),\eqno(2.6)
33~$$
where ${}^rL^{*\vp}$ is a restricted dual module of $L$
defined by ${}^rL^*=\oplus_{\lm}L^{*\vp}_{\lm}$,
$\widehat L$ is a weight completion of $L$ defined by
$\widehat L=\prod_{\lm}L_{\lm}$ and note that
as a $B$-module: $({}^rL^{*\vp})^{*\vpi}\cong \widehat L$.
Similarly, we obtain
\proclaim Corollary 2.2.
For $\bup$-modules  $L$, $M$ and $\uup$-module $N$, there is an isomorphism,
$$
\hom_{\uup}({}^rL^{*\vp}\ot M,N)\cong\hom_{\bup}(M,\widehat L\ot N).
$$
Note that in the rest of this paper the expression
 $L\widehat\ot N$ implies $\widehat L\ot N$.

\vskip2mm
\nd{\sl 2.2. Highest weight $B$-modules\q}
We shall discuss highest weight $B$-modules.

\proclaim Proposition 2.3.
For $\lm\in \ttt^*$, we set
$$
\eqalign{
&H(\lm)=B\big/
\sum_iB{e''_i}+\sum_{h\in P^*}B(q^h-q^{\lan h,\lm\ran}),\cr
&H^r(\lm)=B/\sum_if_iB+\sum_{h\in P^*} (q^h-q^{\lan h,\lm\ran})B.\cr}
 \eqno(2.7)
$$
Then for an arbitrary
 $\lm$, $H(\lm)$ (resp. $H^r(\lm)$)
is an irreducible highest weight left (resp. right) $B$-module
and is a free and rank one $U^-$ (resp. $B^+$)-module.

We denote the highest weight vector $1{\hbox{ mod }}
\sum_iB{e''_i}+\sum_{h\in P^*}B(q^h-q^{\lan h,\lm\ran})$ by $u_{\lm}$ and
$1{\hbox{ mod }}
\sum_if_iB+\sum_{h\in P^*}(q^h-q^{\lan h,\lm\ran})B$ by $u^r_{\lm}$.

\vskip3mm
\nd{\sl Proof.} We show only for $H(\lm)$. In [K1], it is shown that
the subalgebra $U^-\subset U$ has a ${B^{\vee}}$-module structure and it is
isomorphic to an  irreducible ${B^{\vee}}$-module
${B^{\vee}}/\sum_i{B^{\vee}}{e''_i}$. Since
${B^{\vee}}$ is a subalgebra of $B$, $H(\lm)$ is regarded as
a ${B^{\vee}}$-module.
We can easily obtain the following isomorphism
of ${B^{\vee}}$-modules and then of $U^-$-module,
$$
\matrix{
&{B^{\vee}}/\sum_i{B^{\vee}}{e''_i}\cong& U^-&\buildrel\sim\over
\longrightarrow &H(\lm),\cr
&&X&\longmapsto& Xu_{\lm}.\cr}\eqno(2.8)
$$
Hence $H(\lm)$ is irreducible as a ${B^{\vee}}$-module
and then irreducible as a
$B$-module.\hfill{Q.E.D.}

\vskip2mm
Let $\catob$ (resp. $\catobr$) be the category of left (resp. right)
$B$-modules $M$ such that
$M$ has a weight space decomposition and
for any element $u\in M$ there exists $l>0$ such that
$e''_{i_1}e''_{i_2}\cdots e''_{i_l}u=0$
(resp. $uf_{i_1}f_{i_2}\cdots f_{i_l}=0$)
 for any $i_1,i_2,\cdots,i_l\in I$.(See[K1]).

\proclaim Proposition 2.4. (See Remark 3.4.10 [K1])
The category $\catob$ (resp. $\catobr$)
is semi-simple, ({\it i.e.} any object is
a direct sum of simple objects) and for any simple object $M$
there exists $\lm \in \ttt^*$ such that $M\cong H(\lm)$
(resp. $M\cong H^r(\lm)$) as a $B$-module.

\nd{\sl Proof.} We shall show only for $\catob$.
Let $M$ be a simple object of $\catob$ and $v_{\lm}$ be
a highest weight vector of $M$ with a highest weight $\lm$,
where a highest weight vector implies a weight vector annihilated by
any $e''_i$ ($i\in I$). Here we set
$u_{\lm}$ a highest weight vector of $H(\lm)$. We can easily know that a map
$$
\matrix{
\pi\,:&H(\lm)&\longrightarrow& M\cr
&Pu_{\lm}&\longmapsto&Pv_{\lm}\cr}
\qq(P\in B)
$$
is $B$-linear and surjective. The kernel of $\pi$ is a $B$-submodule of
$H(\lm)$, and  by Proposition 2.3, the kernel of $\pi$ is 0.
Hence $\pi$ is injective.
Next we show  the semi-simplicity of
$\catob$. First note that if $N\subset M$ are objects in $\catob$, then
$M/N$ is also an object in $\catob$.
Let $M$ be a non-simple object of $\catob$.
Without a loss of generality, we may assume that $M$ has two highest weight
vectors $u$ and $v$. By the argument in this proof, $Bu$ and $Bv$ are simple.
We have $M=Bu+Bv$ and then $B$-module $M/Bu$ has only
one highest weight vector $\ovl v$ and $M/Bu\cong B\ovl v$.
By the argument in this proof, we have $B\ovl v\cong Bv$ since
$wt(v)=wt(\ovl v)$. Thus  the following exact sequence splits;
$$
0\longrightarrow Bu\longrightarrow M\longrightarrow B\ovl v\longrightarrow0.
$$
Therefore, we obtain the desired result.\hfill{Q.E.D.}

\vskip 3mm
\nd Note that lowest weight $\ovl B$-modules, {\it  e.g.} $H(\lm)^*$ have
similar properties.
\beginsection\S3.  Bilinear forms

In this section, after recalling the Killing form of $U$,
we give an interpretation
of the Killing form of $U$ by the algebra ${B^{\vee}}$.
We also introduce a bilinear pairing similar to a vacuum expectation value.

\proclaim Proposition 3.1. ([R],[T])
(1) There exists a unique bilinear form
$$
(\q,\q):\uup\times \ulow\longrightarrow \FF,\eqno(3.1)
$$
satisfying the following properties;
$$
\eqalignno{
&(x,y_1y_2)=(\Del(x),y_1\ot y_2),\q(x\in \uup,\,y_1,y_2\in \ulow),&(3.2)\cr
&(x_1x_2,y)=(x_2\ot x_1,\Del(y)),\q(x_1,x_2\in \uup,\,y\in \ulow),&(3.3)\cr
&(q^h,q^{h'})=q^{-(h|h')}\q(h,h'\in P^*),&(3.4)\cr
&(T,f_i)=(e_i,T)=0,&(3.5)\cr
&(e_i,f_j)=\del_{ij}/(q_i^{-1}-q_i),&(3.6)\cr}
$$
where $(\,|\,)$ is an invariant bilinear form on $\ttt$.  ([Kac]).\hfill\break
(2) The bilinear form $(\,,\,)$ enjoys the following properties;
$$
\eqalignno{
&(xq^h,yq^{h'})=q^{-(h|h')}(x,y),\q{\hbox{for }}x\in \uup,\,y\in \ulow,
\,h,h'\in P^*,&(3.7)\cr
&{\hbox{For any }}\beta\in Q_+,\,(\,\,,\,\,)_{|U^+_{\beta}\times U^-_{-\beta}}
{\hbox{ is non-degenerate and }}
(U^+_{\gamma},U^-_{-\del})=0,\,{\hbox{ if }}
\gamma\ne \del.&(3.8)\cr}
$$
We call this bilinear form the {\it Killing form} of $U$.

\nd By using the relation (1.5), it is easy to see
that the algebra ${B^{\vee}}$ has the following decomposition;
$$
{B^{\vee}}=\FF\oplus (\sum_if_i{B^{\vee}}+\sum_i{B^{\vee}}e''_i).\eqno(3.9)
$$
Hence for any $x\in {B^{\vee}}$ there is a unique constant $c$ such that
$x\equiv c$ mod $\sum_if_i{B^{\vee}}+\sum_i{B^{\vee}}e''_i$.
We denote this $c$ by $\iota(x)$.

There is the following connection between $\iota$
and the Killing form of $U$.
\proclaim Proposition 3.2.
Let $\iota$ be as above and $(\,,\,)$ the Killing form of $U$.
For any $u\in U^+$ and $v\in U^-$,
$$
\iota(\vp(u)v)=(u,v).\eqno(3.10)
$$

\nd Note that since $u\in U^+=\ovl B^{\vee}\cap U$, $\vp(u)\in B^{\vee}$
and then $\vp(u)v\in {B^{\vee}}$.

\nd{\sl Proof.} We may assume $u$ and $v$ are weight vectors.
If wt$(u)+$wt$(v)\ne0$, trivially $\iota(\vp(u)v)=(u,v)=0$. For
 $u\in U^+_{\beta}$ and $v\in U^-_{-\beta}$ ($\beta\in Q_+$),
it is enough to show
$$
\vp(u)v\equiv (u,v)\,{\hbox{mod}}\,\sum_iBe''_i.\eqno(3.11)
$$
We shall show by the induction on $|\beta|=$ height of $\beta$.
Set $l=|\beta|$. Without a loss of generality, we can set
$u=e_{i_1}e_{i_2}\cdots e_{i_l}$ and $v=f_{j_1}f_{j_2}\cdots f_{j_l}$,
where $\al_{i_1}+\cdots+ \al_{i_l}=\al_{j_1}+\cdots +\al_{j_l}=\beta$.
$$
\eqalign{
&\{\prod_{k=1}^l(q_{i_k}^{-1}-q_{i_k})\}\vp(u)v
=e''_{i_l}\cdots e''_{i_2} e''_{i_1}f_{j_1}f_{j_2}\cdots f_{j_l}\cr
&=q_{i_1}^{\lan h_{i_1},\al_{j_1}+\cdots+\al_{j_l}\ran}
e''_{i_l}\cdots e''_{i_2}f_{j_1}\cdots f_{j_l}e''_{i_1}\cr
&\qq\qq
+\sum_{m=1}^l q_{i_1}^{\lan h_{i_1},\al_{j_1}+\cdots+\al_{j_{m-1}}\ran}
\del_{i_1,j_m}e''_{i_l}\cdots e''_{i_2}f_{j_1}\cdots f_{j_{m-1}}
f_{j_{m+1}}\cdots f_{j_l}.\cr}
$$
Thus, by the hypothesis of the induction,
$$
\eqalign{
&\vp(u)v\cr
&\equiv
{1\over{(q_{i_1}^{-1}-q_{i_1})}}
\sum_{m=1}^l q_{i_1}^{\lan h_{i_1},\al_{j_1}+\cdot\cdot+\al_{j_{m-1}}\ran}
\del_{i_1,j_m}\vp(e_{i_2}\cdot\cdot e_{i_l})f_{j_1}\cdot\cdot f_{j_{m-1}}
f_{j_{m+1}}\cdot\cdot f_{j_l}\,{\hbox{mod}}\,\sum_iBe''_i\cr
&\equiv{1\over{(q_{i_1}^{-1}-q_{i_1})}}
\sum_{m=1}^l q_{i_1}^{\lan h_{i_1},\al_{j_1}+\cdot\cdot+\al_{j_{m-1}}\ran}
\del_{i_1,j_m}(e_{i_2}\cdot\cdot e_{i_l},f_{j_1}\cdot\cdot f_{j_{m-1}}
f_{j_{m+1}}\cdot\cdot f_{j_l})\,{\hbox{mod}}\,\sum_iBe''_i.\cr}\eqno(3.12)
$$
On the other hand, from the formulae (3.2)--(3.8)
and the explicit form of $\Del(f_i)$.
$$
\eqalignno{
&(e_{i_1}\cdots e_{i_l},f_{j_1}\cdots f_{j_l})
=(e_{i_2}\cdots e_{i_l}\ot e_{i_1},\Del(f_{j_1}\cdots f_{j_l}))\cr
&=\sum_{m=1}^l(e_{i_2}\cdots e_{i_l}\ot e_{i_1},
f_{j_1}\cdots f_{j_{m-1}}f_{j_{m+1}}\cdots f_{j_l}\ot
t^{-1}_{j_1}\cdots t^{-1}_{j_{m-1}}f_{j_m}t^{-1}_{j_{m+1}}\cdots
t^{-1}_{j_l})\cr
&=\sum_{m=1}^l(e_{i_2}\cdots e_{i_l},
f_{j_1}\cdots f_{j_{m-1}}f_{j_{m+1}}\cdots f_{j_l})
(e_{i_1},t^{-1}_{j_1}\cdots t^{-1}_{j_{m-1}}f_{j_m}t^{-1}_{j_{m+1}}
\cdots t^{-1}_{j_l})\cr
&=\sum_{m=1}^l q_{j_m}^{\lan h_{j_1}+\cdots+h_{j_{m-1}},\al_{j_m}\ran}
(e_{i_1},f_{j_m})(e_{i_2}\cdots e_{i_l},
f_{j_1}\cdots f_{j_{m-1}}f_{j_{m+1}}\cdots f_{j_l}) \cr
&=\sum_{m=1}^l
{{q_{i_1}^{\lan h_{i_1},\al_{j_1}+\cdots+\al_{j_{m-1}}\ran}\del_{i_1,j_m}}
\over{(q_{i_1}^{-1}-q_{i_1})}}
(e_{i_2}\cdots e_{i_l},
f_{j_1}\cdots f_{j_{m-1}}f_{j_{m+1}}\cdots f_{j_l}).&(3.13) \cr}
$$
{}From the equality of
(3.12) and (3.13), we get the desired result. \hfill{Q.E.D.}

\vskip3mm
We shall define a bilinear pairing
similar to vacuum expectation values.
For $\lm\in \ttt^*_{cl}$ we define a bilinear pairing
$\lan\q|\q\ran: H^r(\lm)\times H(\lm)\lar \FF$ as follows:
Similar to (3.9) the algebra $B$  has a decomposition;
$$
B=T\oplus(\sum_i f_iB+\sum_i Be''_i).\eqno(3.14)
$$
Let $\Omega:B\lar T$ be a canonical projection. Here we can define
a $T$-valued pairing $E:B\times B\lar T$ by
$E(x,y)=\Omega(xy)$ for $x,y\in B$. By the definition of $E(\,,\,)$  and
the associativity of $B$,  we have
$$
E(xy,z)=E(x,yz)\qq{\hbox{ for }}x,y,z\in B.\eqno(3.15)
$$
We define $\pi_{\lm}:T\lar \FF$ by $tu_{\lm}=\pi_{\lm}(t)u_{\lm}$ for $t\in T$.
A bilinear pairing
$\lan\q|\q\ran: H^r(\lm)\times H(\lm)\lar \FF$ is given by
$\lan u|v\ran=\pi_{\lm}(E(P,Q)),$ where $u=u^r_{\lm}P$ and $v=Qu_{\lm}$
($P,Q\in B$). It is clear that
this is well-deined, {\it i.e.} it does not depend on a
choice of $P$ and $Q$.

\proclaim Proposition 3.3.
There is a  unique and non-degenerate
bilinear pairing $\lan\q|\q\ran: H^r(\lm)\times H(\lm)\lar \FF$
such that
$$
\lan ux|v\ran =\lan u|xv\ran,\,(x\in B)\q
\hbox{and}\q \lan u^r_{\lm}|u_{\lm}\ran=1.\eqno(3.16)
$$

\nd{\sl Proof.}
If we assume the existence, then the uniqueness immediately
follows from (3.16).
The existence follows from the construction above and (3.15).
We shall show non-degeneracy.
Let $\{P_i\}\subset U^+$ and $\{Q_i\}\subset U^-$ be bases
dual to each other with respect to the Killing form such that each basis
element is a weight vector.
By Proposition 3.2, we get
$$
\vp(P_i)Q_j\equiv \del_{i,j}\,{\hbox{mod}}\,
\sum_if_i{B^{\vee}}+\sum_i{B^{\vee}}e''_i.
\eqno(3.17)
$$
Hence
$$
\lan u^r_{\lm}\vp(P_i)|Q_ju_{\lm}\ran=\del_{i,j}.
$$
Moreover, by Proposition 2.3, $\{u^r_{\lm}\vp(P_i)\}$ and $\{Q_iu_{\lm}\}$
are bases of $H^r(\lm)$ and $H(\lm)$ respectively.
Thus we have completed the proof of Proposition 3.3.\hfill{Q.E.D.}
\par
\vskip2mm
\nd From the property (3.16),
we shall use the expression $\lan u|x|v\ran$ for
$\lan ux|v\ran=\lan u|xv\ran$ ($u\in H^r(\lm)$, $v\in H(\lm)$
and $x\in B$).

\beginsection\S4. Intertwiners

In this section and the next section,
 we restrict $\ge$ to be an affine Lie algebra.
We shall study the following type of intertwiners,
 which is an analogue of so-called
``$q$-vertex operators''([FR],[DJO]).
$$
\hom_{B_q(\ge)}(H(\lm), H(\mu)\widehat\ot V_z),\eqno(4.1)
$$
where $V_z$ is a representation of $U=\uq$ (see below).

\nd{\sl 4.1. Notations \q}
We shall prepare notations. (See [KMN${}^2$],[DJO].)

\nd Set $I=\{0,1,\cdots,n\}$ and
$(\lan h_i,\al_j\ran)_{0\leq i,j\leq n}$ coincides
with an affine Cartan matrix in [Kac] except for the type $A_{2n}^{(2)}$.
For this type we reverse the ordering of vertices since we need that
$\del-\al_0\in \sum_{i=1}^n\ZZ\al_i$ for a generator of null roots $\del$.
Let  $c$ be a canonical center of $\ge$,
$\{\Lm_i\}_{i\in I}$  a set
of fundamental weights and $d\in\ttt$  a scaling element.
 Now, since $\ge$ is affine, dim$\ttt=\#I+1$. Thus we can write
$P=\oplus_i\ZZ\Lm_i\oplus\ZZ\del$ and $P^*=\oplus_i\ZZ h_i\oplus\ZZ d$.
We set $\ttt^*_{cl}=\oplus_i\QQ\Lm_i$,
$P_{cl}=P/\ZZ\del$, $(P_{cl})^*=\oplus_{i=0}^n\ZZ h_i$ and let
$cl:P\rightarrow P_{cl}$ be a canonical projection. We fix a map
$af:P_{cl}\rightarrow P$ by $af\circ cl(\al_i)=\al_i$ $(i\ne 0)$ and
$af\circ cl(\Lm_0)=\Lm_0$ so that $cl\circ af=$id and
$af\circ cl(\al_0)=\al_0-\del$.
For a fixed $k\in \QQ$, we set
$(\ttt_{cl}^{*})_k=\{\lm\in \ttt^*_{cl}|\lan c,\lm\ran=k\}$ and
we say that $\lm\in (\ttt^{*}_{cl})_k$ has a {\it level} $k$.
The subalgebra of $U$ (resp. B) generated by
$\{e_i\,({\hbox{resp. }}e''_i),f_i|i\in I\}$
and $q^h$ $(h\in (P_{cl})^*)$ is denoted by $U'$ (resp. $B'$).

For a finite dimensional $U'$-module $V$ and a formal variable $z$,
we define an affinization
$V_z=\FF[z,z^{-1}]\ot V$ with a $U$-module structure as follows;
$$
\eqalign{
&e_i(z^n\ot u)=z^{n+\del_{i0}}\ot e_iu,\qq
f_i(z^n\ot u)=z^{n-\del_{i0}}\ot f_iu,\cr
& {\hbox{wt}}(z^n\ot u)=n\del+af({\hbox{wt }}u).\cr}\eqno(4.2)
$$

\nd{\sl 4.2. Condition for existence \q}
We shall examine the condition for
existence  of the intertwiners of $B$-modules
of type (4.1) by the similar way of [DJO].

\proclaim Definition 4.1.
For $\lm$, $\mu\in (\ttt^{*}_{cl})_k$  and
$\Phi\in\hom_{B}(H(\lm), H(\mu)\widehat\ot V_z)$ and the
highest weight vector $u_{\lm}$ and $u_{\mu}$,
write the image of $u_{\lm}$ by $\Phi$
$$
\Phi u_{\lm}=u_{\mu}\ot v_{lt}+\cdots,
$$
where $\cdots$ implies terms of the form $u\ot v$ with
$u\in\oplus_{\xi\ne\mu}H(\mu)_{\xi}$. We call $v_{lt}$
the {\it leading term} of $\Phi$.

\proclaim Proposition 4.2.
The map sending $\Phi$ to its leading term gives an isomorphism;
$$
\hom_{B}(H(\lm), H(\mu)\widehat\ot V_z)\buildrel\sim\over\longrightarrow
(V_z)_{\lm-\mu}.
$$

\nd{\sl Proof.} Let $\FF u_{\lm}$ be one dimensional $\bup$-module with
defining relations: $e''_iu_{\lm}=0$ and
$q^hu_{\lm}=q^{\lan h,\lm\ran}u_{\lm}$. We prepare the following lemma.

\proclaim Lemma 4.3.
We have the following isomorphism;
$$
\matrix{
&\hom_{B}(H(\lm), H(\mu)\widehat\ot V_z)
&\buildrel\sim\over\longrightarrow
&\hom_{\bup}(\FF u_{\lm}, H(\mu)\widehat\ot V_z)\cr
&\Phi&\longmapsto&\Phi_{|\FF u_{\lm}}\cr}\eqno(4.3)
$$

\nd{\sl Proof of Lemma 4.3.} By $B$-linearity of $\Phi$,
one gets $\bup$-linearity of $\Phi_{|\FF u_{\lm}}$  and
if $\Phi u_{\lm}=0$, then $\Phi=0$. Hence the map (4.3) is well-defined
and injective. To show the surjectivity,
take a vector $v\in  H(\mu)\widehat\ot V_z$
such that wt$(v)=\lm$ and $e''_iv=0$
for all $i\in I$. By the property of
the category $\catob$ (Proposition 2.4.), $B$-module $Bv$ is isomorphic to
$H(\lm)$ as a $B$-module.
Hence we obtain the surjectivity. \hfill{Q.E.D.}

\vskip3mm
{}From Corollary 2.2, we have the following
isomorphism;
$$
{\hbox{R.H.S. of }}(4.3)\cong
\hom_{\uup}({}^rH(\mu)^{*\vp}\ot \FF u_{\lm},V_z).\eqno(4.4)
$$
Here note that $\Delb(\uup)\subset \uup\ot \bup$ and
as a $\uup(=\ovl B\cap U)$-module
${}^rH(\mu)^{*\vp}$ is isomorphic to
$$
\uup/\sum_{h\in P^*}\uup(q^h-q^{-\lan h,\mu\ran}).
$$
It is easy to see that R.H.S of (4.4) is isomorphic to $(V_z)_{\lm-\mu}$.
\hfill{Q.E.D.}
\beginsection\S5. 2-point functions and commutation relations of intertwiners

In this section we show
that a matrix determined by ''2-point functions''  coincides with
a quantum R-matrix up to a diagonal matrix
and give commutation relations for intertwiners.

\vskip 2mm
\nd{\sl 5.1. 2-point functions\q}
First we shall define ``2-point functions'' for
the intertwiners of $B$-modules introduced in Sect.4.
We fix $k\in \QQ$. For $\Phi_{\lm}^{\mu\,V}(z_1)\in
\hom_{B}(H(\lm), H(\mu)\widehat\ot V_{z_1})$ and
$\Phi_{\mu}^{\nu\,W}(z_2)\in
\hom_{B}(H(\mu), H(\nu)\widehat\ot W_{z_2})$
($\lm,\mu,\nu\in (\ttt^*_{cl})_k$),
we use an  abbreviated notation
 $\Phi_{\mu}^{\nu\,W}(z_2)\Phi_{\lm}^{\mu\,V}(z_1)$
for
$(\Phi_{\mu}^{\nu\,W}(z_2)\ot{\hbox{id}}_{V_{z_1}})\Phi_{\lm}^{\mu\,V}(z_1)$.
With this notation, the following is called {\it 2-point function;}
$$
\lan u^r_{\nu}|
\Phi_{\mu}^{\nu\,W}(z_2)\Phi_{\lm}^{\mu\,V}(z_1)|u_{\lm}\ran
\in \FF[[{{z_1}\over{z_2}}]]\ot W\ot V,
$$

We shall give an explicit description of 2-point functions.
For a $B$-module $H(\lm)$,
${}^rH(\lm)^{*\vp}$  means the restricted dual module
$\oplus_{\xi}(H(\lm)^*)_{\xi}$ as in Sect.2.
Here ${}^rH(\lm)^{*\vp}$ is an irreducible lowest weight left $\ovl B$-module
with a lowest weight vector denoted by $u^*_{\lm}$ such that
$f'_iu^*_{\lm}=0$ for any $i\in I$,
$q^h=q^{-\lan h,\lm\ran}u^*_{\lm}$ for any $h\in P^*$, $(u^*_{\lm},u_{\lm})=1$
and $(u^*_{\lm},v)=0$ for $v\in \oplus_{\mu\ne\lm}H(\lm)_{\mu}$.
{}From Proposition 2.1 and the formula (2.6),
there is an isomorphism for $\lm,\,\mu\in (\ttt^*_{cl})_k$;
$$
\Psi:\hom_U({}^rH(\mu)^{*\vp}\ot H(\lm),V_z)\buildrel\sim\over\longrightarrow
\hom_B(H(\lm),H(\mu)\widehat\ot V_z).\eqno(5.1)
$$
We translate this in terms of dual bases as follows.
Let $\{u_i\}\subset H(\mu)$ and $\{u_i^*\}\subset {}^rH(\mu)^{*\vp}$ be bases
dual to each other such that $u_{\mu}\in\{u_i\}$.
Then for $x\in H(\lm)$ and $\phi\in\hom_U({}^rH(\mu)^{*\vp}\ot H(\lm),V_z)$,
$\Psi$ is given by
$$
\Psi(\phi)(x)=\sum_iu_i\ot \phi(u_i^*\ot x).\eqno(5.2)
$$
The following lemma is immediate from (5.2) and the definition of the
leading term.
\proclaim Lemma 5.1.
Let $\Psi$ and $\phi$ be as above.
Then $\phi(u^*_{\mu}\ot u_{\lm})$ is a leading term of $\Psi(\phi)$.

\proclaim Lemma 5.2.
Let $\{P_i\}\subset U^+$ and $\{Q_i\}\subset U^-$ be bases dual to each other
with respect to the Killing form such that each basis element is
a weight vector and $1\in \{P_i\}$ (and then
$1\in \{Q_i\}$).
Then for any $\lm\in \ttt^*$, $\{P_iu^*_{\lm}\}\subset {}^rH(\lm)^{*\vp}$ and
$\{Q_iu_{\lm}\}\subset H(\lm)$ are bases dual to each other.

\nd{\sl  Proof.} First note that for $u\in {}^rH(\lm)^{*\vp}$,
$v\in H(\lm)$, $x\in \ovl B$  and $y\in B$,
$$
(xu,yv)=(u,\vp(x)yv)=(\vpi(y)xu,v).\eqno(5.3)
$$
{}From Proposition 3.2 and (5.3),
we get $(P_iu^*_{\lm},Q_ju_{\lm})=\del_{i,j}$ and
from Proposition 2.3 (and similar one for lowest $\ovl B$-modules),
we know that $\{P_iu^*_{\lm}\}$ and $\{Q_iu_{\lm}\}$ are bases.
\hfill{Q.E.D.}

\vskip3mm

Let $\UR$ be a universal R-matrix and $\UR'(z)$  a modified universal
R-matrix as in (B10) (see Appendix B.).
Let $V$ and $W$ be finite dimensional $U'$-modules and
 $V_{z_1}$ and $W_{z_2}$  their affinizations.
We denote the image of the universal R-matrix
 onto a $U$-module $V_{z_1}\ot W_{z_2}$
by $R^{VW}(z)=\pi_{V\ot W}(\UR'(z))$, where $z=z_1/z_2$. This coincides
with a quantum R-matrix on $V\ot W$ up to a scalar factor.

\vskip3mm
\proclaim Theorem 5.3.
For intertwiners $\Phi_{\lm}^{\mu V}(z_1)\in
\hom_B(H(\lm),H(\mu)\widehat\ot V_{z_1})$ and $\Phi_{\mu}^{\nu W}(z_2)
\in \hom_B(H(\mu),H(\nu)\widehat\ot W_{z_2})$, we set $v_0\in V_{z_1}$ and
$w_0\in W_{z_2}$ be leading terms of $\Phi_{\lm}^{\mu V}(z_1)$ and
$\Phi_{\mu}^{\nu W}(z_2)$ respectively. Then the 2-point function is given by
$$
\lan u^r_{\nu}|\Phi_{\mu}^{\nu\,W}(z_2)\Phi_{\lm}^{\mu\,V}(z_1)|u_{\lm}\ran
=q^{(\lm-\mu,\mu-\nu)}\sigma\circ R^{VW}(z_1/z_2)(v_0\ot w_0),
$$
where $\sigma:a\ot b\lar b\ot a$.

\vskip3mm
\nd{\sl Proof.}  Let $\Psi$ be as in (5.1).
We set
$\phi_1=\Psi^{-1}(\Phi_{\lm}^{\mu V}(z_1))$ and
$\phi_2=\Psi^{-1}(\Phi_{\mu}^{\nu W}(z_2))$.
Let $\{P_i\}$ and $\{Q_i\}$ be as in Lemma 5.2. From (5.2)
and Lemma 5.2, for $x\in H(\lm)$ we have
$$
\Phi_{\mu}^{\nu W}(z_2)\Phi_{\lm}^{\mu V}(z_1)(x)
=\sum_{i,j}Q_ju_{\nu}\ot \phi_2(P_ju^*_{\nu}\ot Q_iu_{\mu})\ot
\phi_1(P_iu^*_{\mu}\ot x),
$$
and then 2-point function can be written by
$$
\lan u^r_{\nu}|\Phi_{\mu}^{\nu W}(z_2)\Phi_{\lm}^{\mu V}(z_1)(x)|u_{\lm}\ran=
\sum_i\phi_2(u^*_{\nu}\ot Q_iu_{\mu})
\ot \phi_1(P_iu^*_{\mu}\ot u_{\lm})\in \FF[[{{z_1}\over{z_2}}]]\ot W\ot V.
\eqno(5.4)
$$
By the intertwining property of $\phi_i$ ($i=1,2$) and the fact
that $e''_iu_{\lm}=0$ and $f'_iu^*_{\mu}$ for any $i \in I$, we have
$$
\eqalign{
&P_i\phi_1(u^*_{\mu}\ot u_{\lm})=\phi_1(\Delb(P_i)(u^*_{\mu}\ot u_{\lm}))
=\phi_1(P_iu^*_{\mu}\ot u_{\lm}),\cr
&Q_i\phi_2(u^*_{\nu}\ot u_{\mu})=\phi_2(\Delb(Q_i)(u^*_{\nu}\ot u_{\mu}))
=\phi_2(u^*_{\nu}\ot Q_iu_{\mu}).\cr}
$$
 Hence (5.4) can be rewritten by
$$
\lan u^r_{\nu}|\Phi_{\mu}^{\nu W}(z_2)\Phi_{\lm}^{\mu V}(z_1)(x)|u_{\lm}\ran=
\sigma\,(\sum_iP_i\ot Q_i)\cdot\{\phi_1(u^*_{\mu}\ot u_{\lm})
\ot\phi_2(u^*_{\nu}\ot u_{\mu})\}.\eqno(5.5)
$$
{}From (B9) in Appendix B, on a vector
$u\ot v$ (wt$(u)=\xi$ and wt$(v)=\eta$) we have
$$
\UR=q^{-(\xi,\eta)}\sum_iP_i\ot Q_i. \eqno(5.6)
$$
{}From Lemma 5.1, $\phi_1(u^*_{\mu}\ot u_{\lm})
\ot\phi_2(u^*_{\nu}\ot u_{\mu})=v_0\ot w_0$.
Therefore by the formulae (5.5) and (5.6), we obtain the desired result.
\hfill{Q.E.D.}

Fix bases $C$ and $C'$  of
$V$ and $W$ respectively such that each basis element is a weight vector.
For a pair $(v_i,w_j)\in C\times C'$
let $\Phi_{\lm,(i)}^{\mu_i\,V}(z_1)
\in \hom_B(H(\lm),H(\mu_i)\widehat\ot V_{z_1})$ and
 $\Phi_{\mu_i,(j)}^{\nu_j\,V}(z_2)
\in \hom_B(H(\mu_i),H(\nu_j)\widehat\ot W_{z_2})$ be intertwiners with leading
terms $v_i$ and $w_j$ respectively. Let $\Xi(z_1,z_2),\,
D\in {\hbox{End}}(V\ot W)$
be matrices defined by
$$
\eqalign{
&\Xi(z_1,z_2):v_i\ot w_j\mapsto \sigma
\lan u^r_{\nu_j}|\Phi_{\mu_i,(j)}^{\nu_j\,W}(z_2)
\Phi_{\lm,(i)}^{\mu_i\,V}(z_1)|u_{\lm}\ran,\cr
&D:v_i\ot w_j\mapsto q^{({\hbox{wt}}(v_i),{\hbox{wt}}(w_j))}v_i\ot w_j.\cr}
$$
{}From Theorem 5.3, we obtain the following;
\proclaim Corollary 5.4. With the notations as above, we have
$$\Xi(z_1,z_2)=DR^{VW}(z_1/z_2).$$

\vskip2mm
\nd{\sl 5.2. Commutation relations\q}
Let $V$ and $W$ be finite dimensional $U'$-modules.
We assume that $V_{z_1}\ot W_{z_2}$ is an irreducible $U$-module.
Let $C$ and $C'$ be bases of $V$ and $W$ as in 5.1.
Now, we fix  $v_0\in C$, $w_0\in C'$, $\lm,\nu\in(\ttt^*_{cl})_k$ such that
$\lm-\nu=af({\hbox{wt}}(v_0)+{\hbox{wt}}(w_0))$
 and let
$\Phi_{\mu}^{\nu\,V}(z)$ and $\Phi_{\lm}^{\mu\,W}(z)$ be intertwiners
such that their leading terms are $v_0\in C$ and
$w_0\in C'$ respectively. Here note that we identify  $v\in V$ and
$w\in W$ with $1\ot v\in V_z$ and $1\ot w\in W_z$ respectively.
We set
$$
E=\{(v,w)\in C\times C'|af({\hbox{wt}}(v))+af({\hbox{wt}}(w))=
af({\hbox{wt}}(v_0))+af({\hbox{wt}}(w_0))\}.
$$
For a pair $(v_i,w_i)\in E$, we set $\Phi_{\lm,(i)}^{\mu_i V}(z)$
and $\Phi_{\mu_i,(i)}^{\nu\,W}(z)$
be intertwiners such that their leading terms
are $v_i$ and $w_i$ respectively.

For a $U'$-modules $V\ot W$,
from the uniqueness and the unitarity of  quantum R-matrices,
there exists $f(x)\in \FF[[x,x^{-1}]]$
such that
$$
\eqalignno{
&R^{VW}(z_1/z_2)\sigma R^{WV}(z_2/z_1)\sigma=f(z_1/z_2){\hbox{id}}_{V\ot W}.&
(5.7)\cr
}
$$
We define $W_i(z_1/z_2)$ by,
$$
R^{VW}(z_1/z_2)^{-1}(v_0\ot w_0)=\sum_i
q^{({\hbox{wt}}(v_i),{\hbox{wt}}(w_i))}
(v_i\ot w_i) W_i(z_1/z_2).\eqno(5.8)
$$

\proclaim Proposition 5.5.
With the notations as above, we have the following commutation relation
(in the sense of matrix element);
$$
\sigma\circ R^{VW}(z_1/z_2)\Phi_{\mu}^{\nu \, V}(z_1)\Phi_{\lm}^{\mu\, W}(z_2)
=q^{(\lm-\mu,\mu-\nu)}
f(z_1/z_2)\sum_i \Phi_{\mu_i,(i)}^{\nu \,W}(z_2)
\Phi_{\lm,(i)}^{\mu_i\,V}(z_1)
W_i(z_1/z_2).
$$

\vskip2mm
\nd{\sl Proof.}
{}From (5.7) and Theorem 5.3, we have
$$
\eqalign{
f(z_1/z_2)(v_0\ot w_0)&=R^{VW}(z_1/z_2)\sigma R^{WV}(z_2/z_1)\sigma
(v_0\ot w_0)\cr
&=q^{-(\lm-\mu,\mu-\nu)}R^{VW}(z_1/z_2)
\lan u^r_{\nu}|\Phi_{\mu}^{\nu \,V}(z_1)
\Phi_{\lm}^{\mu\,W}(z_2)|u_{\lm}\ran.}\eqno(5.9)
$$
On the other hand, from (5.8) and Theorem 5.3,
$$
v_0\ot w_0=\sigma\sum_i\lan u^r_{\nu}|
\Phi_{\mu_i,(i)}^{\nu \,W}(z_2)
\Phi_{\lm,(i)}^{\mu_i\,V}(z_1)|u_{\lm} \ran W_i(z_1/z_2).
\eqno(5.10)
$$
{}From (5.9), (5.10), the intertwining property
of $\sigma\circ R^{VW}(z)$
and $B$-linearity of elements in $\hom_B(H(\lm),H(\mu)\widehat\ot V_z)$,
 we obtain the desired result.
\hfill{Q.E.D.}

\vskip 3mm
\nd{\bf Example.}
Set $\ge=\widehat{\ssl}_2$ and
$V=\FF u_+ \oplus\FF u_-$. A $U$-module structure of $V_z$
is given by
$$
\eqalign{
&e_0(z^nu_+)=z^{n+1}u_-,\,e_0(z^nu_-)=0,\,f_0(z^nu_+)=0,\,
f_0(z^nu_-)=z^{n-1}u_+,\cr
&e_1(z^nu_+)=0,\,e_1(z^nu_-)=z^nu_+,\,f_1(z^nu_+)=z^nu_-,\,f_1(z^nu_-)=0,\,\cr
&{\hbox{wt}}(z^nu_{\pm})=n\del\pm(\Lm_1-\Lm_0).\cr}
$$
Set
$$
(z)_{\infty}=\prod_{i=0}^{\infty}(1-q^{4i}z), \qq
\rho(z)={{(q^2z)_{\infty}^2}\over{(z)_{\infty}(q^4z)_{\infty}}},
\qq \Theta(z)=(z)_{\infty}(q^4z^{-1})_{\infty}(q^4)_{\infty}.
$$
An explicit form of the image of the universal R-matrix
onto $V_{z_1}\ot V_{z_2}$ is described in  [DFJMN], therefore
2-point functions are given as follows;
$$
\eqalign{
&\lan u^r_{\nu}|\Phi_{\mu}^{\nu V}(z_2)\Phi_{\lm}^{\mu V}(z_1)|u_{\lm}\ran
=\rho(z_1/z_2)\times\cr
&\cases{
u_{\pm}\ot u_{\pm}& if $\lm-\mu=\mu-\nu=\pm(\Lm_1-\Lm_0)$,\cr
{{q^{-1}-q}\over{1-q^2z_1/z_2}}{{z_1}\over{z_2}}u_+\ot u_-+
{{1-z_1/z_2}\over{1-q^2z_1/z_2}}u_-\ot u_+
& if $\lm-\mu=\nu-\mu=\Lm_1-\Lm_0$,\cr
{{1-z_1/z_2}\over{1-q^2z_1/z_2}}u_+\ot u_-
+{{q^{-1}-q}\over{1-q^2z_1/z_2}}u_-\ot u_+
& if $\lm-\mu=\nu-\mu=\Lm_0-\Lm_1$,\cr}\cr}
$$
where we normalize intertwiners so that their leading term is $u_+$ or $u_-$.
Note that $(\Lm_i,\Lm_j)=\del_{i1}\del_{j1}/2$.
The function in (5.7) is given by
$$
f(z)=q^{-1}{{\Theta(q^2z)^2}\over{\Theta(z)\Theta(q^4z)}}.
$$
\beginsection\S6.  An element $\widetilde\UR$ and a projector $\Gamma$

In this section we do not restrict $\ge$ to be an affine Lie algebra.
We introduce an element $\wtil\UR$,
which satisfies the properties similar to those of the universal R-matrix.

\vskip2mm
\nd{\sl 6.1. An element $\wtil\UR$\q}
We follow the notations as in Appendix B. We can define
$(\widehat B\widehat\ot\widehat U^{\widehat\ot n})^{\widehat{}}$ and
extend $\Delr\ot 1^{\ot n}$ by the similar manner as in Appendix B.

Let $\UR$ be the universal R-matrix of $U$ (see (B8) in Appendix B.).
We define
$$
\wtil\UR
=q^{-H}\sum_{\beta\in Q_+}q^{(\beta,\beta)}
(k_{\beta}^{-1}\ot k_{\beta})(\vp S^{-1}\ot1)(C_{\beta})
\in ((\widehat B\widehat\ot\widehat U)^{\widehat{}}.\eqno(6.1)
$$
Here note that
left components of $C_{\beta}$
belongs to $\uup$, then
the map $\vp S^{-1}: \uup\buildrel{S^{-1}}\over\lar
 \uup
\buildrel{\vp}\over\lar \bup$ is well-defined.
and formally we can write $\wtil\UR=(\vp S^{-1}\ot1)\UR$ since
$\vp S^{-1}$ act as an identity for the Cartan part.

\vskip3mm
\proclaim Proposition 6.1.
$\wtil\UR$ enjoys the following properties;
$$
\eqalignno{
&\wtil\UR {\hbox { is invertible and }}\cr
&\wtil\UR^{-1}
=\sum_{\beta\in Q_+}q^{(\beta,\beta)}(1\ot k_{\beta})(\vp\ot 1)(C_{\beta})
q^H,&(6.2)\cr
&(\Delr\ot1)\wtil\UR=\wtil\UR_{13}\UR_{23},&(6.3)\cr
&(1\ot \Del)\wtil\UR=\wtil\UR_{13}\wtil\UR_{12},&(6.4)\cr
&\wtil\UR\cdot\Delr(X)=(X\ot1)\cdot\wtil\UR \,\q(X\in U^-),&(6.5)\cr
&\wtil\UR\cdot(\vp\ot S)\sigma\Delr(X)=(\vp S^{-1}\ot S\vpi)
\sigma\Delr(X)\cdot\wtil\UR,\q(X\in B^+).
&(6.6)\cr}
$$

\proclaim Corollary 6.2.
We have the following equation in $(\widehat B\widehat\ot\widehat
U\widehat\ot\widehat U)^{\widehat{}}$;
$$
\UR_{23}\wtil\UR_{13}\wtil\UR_{12}=\wtil\UR_{12}\wtil\UR_{13}\UR_{23}.
$$

\nd{\sl Proof of Corollary 6.2.} From the properties (6.4) and (B1),
$$
\eqalignno{
\UR_{23}\wtil\UR_{13}\wtil\UR_{12}&=\UR_{23}(1\ot\Del)\wtil\UR&{}\cr
&=(1\ot\sigma\circ\Del)\wtil\UR\cdot\UR_{23}&{}\cr
&=\wtil\UR_{12}\wtil\UR_{13}\UR_{23}.&{\hbox{Q.E.D.}}\cr}
$$

\nd{\sl Proof of Proposition 6.1.}
We can derive (6.2), (6.3), (6.4) and (6.6)
from the property of
$\UR$. In fact, (6.2), (6.4) and (6.6) are immediate from (B1)--(B3).
To show (6.3), we only need the following
$$
\Delr(\vp S^{-1}(X))=(\vp S^{-1}\ot1)\Del(X),\q{\hbox{ for any }}X\in \uup.
$$
This is easily obtained by direct calculations.
Hence
$$
\eqalign{
&(\Delr\ot1)\wtil\UR=(\Delr\ot1)(\vp S^{-1}\ot1)\UR\cr
&=(\vp S^{-1}\ot1\ot1)(\Del\ot1)\UR=(\vp S^{-1}\ot1\ot1)\UR_{13}\UR_{23}
=\wtil\UR_{13}\UR_{23}.\cr}
$$
In order to show (6.5), we shall prepare some lemmas.
\proclaim Lemma 6.3.
Let $C_{\beta}$ be as in Appendix B.
Set $\wtil C_{\beta}=(\vp S^{-1}\ot1)C_{\beta}$.
For any $i\in I$ we have,
$$
[f_i\ot1,\wtil C_{\beta+\al_i}]=\wtil C_{\beta}(t_i\ot f_i).\eqno(6.7)
$$

\vskip 2mm
\nd{\sl Proof.} We show the following lemma.
\vskip 2mm
\proclaim Lemma 6.4.
For any $i\in I$, $\beta\in Q_+$ and $u\in U^+_{\beta+\al_i}$, we have
$$
[f_i,\vp S^{-1}(u)]={{\vp S^{-1}(v)t_i}\over{\qii-\qi}},
$$
where $v\in U^+_{\beta}$ is uniquely determined
 by $\Del(u)=u\ot1+vt_i\ot e_i+\cdots$, where $\cdots$ implies
terms whose right component is an element of
$\oplus_{\beta\ne0,\al_i}U^+_{\beta}$

\nd{\sl Proof.} We fix $i\in I$. For $\beta=\sum_jm_j\al_j$, we shall
show the induction on $m_i$. We may assume that $u$ is a monomial
$e_{j_1}e_{j_2}\cdots e_{j_l}$ where $l=|\beta|+1$.
If $m_i=0$, we can write $u=ze_iw$ where $z=e_{j_1}\cdots e_{j_{k-1}}$,
$w=e_{j_{k+1}}\cdots e_{j_l}$ and $j_p\ne i$
for any $p=1,\cdots,k-1,k+1,\cdots,l$.
It is easy to see
$$
[f_i,\vp S^{-1}(e_j)]={{\del_{i,j}t_i}\over{\qii-\qi}}.\eqno(6.8)
$$
{}From (6.8), we get
$[f_i,\vp S^{-1}(z)]=[f_i,\vp S^{-1}(w)]=0$.
Thus $[f_i,\vp S^{-1}(u)]={{\vp S^{-1}(zt_iw)}\over{\qii-\qi}}$.
On the other hand,
$$
\eqalign{
\Del(ze_iw)=\Del(z)\Del(e_i)\Del(w)
&=(z\ot1+\cdots)(e_i\ot1+t_i\ot e_i)(w\ot1+\cdots)\cr
&=ze_iw\ot1+zt_iw\ot e_i+\cdots.\cr}
$$
Therefore we get $vt_i=zt_iw$. Hence, we have
$\vp S^{-1}(v)t_i=\vp S^{-1}(vt_i)=\vp S^{-1}(zt_iw)$.
The case of $m_i=0$ is completed.

For $m_i>0$, we can write $u=u'u''$  such that $0<m'_i<m_i$ and $0<m''_i<m_i$
where $m'_i$ and $m''_i$ are given by
wt$(u')=\sum_jm'_j\al_j$ and wt$(u'')=\sum_jm''_j\al_j$. Let
$v'$ and $v''$ be vectors given by
$$
\Del(u')=u'\ot1+v't_i\ot e_i+\cdots,\,{\hbox{ and }}\,
\Del(u'')=u''\ot 1+v''t_i\ot e_i+\cdots.
$$
Then we obtain
$$
\eqalign{
\Del(u)=\Del(u')\Del(u'')&=
(u'\ot1+v't_i\ot e_i+\cdots)(u''\ot 1+v''t_i\ot e_i+\cdots)\cr
&=u'u''\ot1+(u'v''t_i+v't_iu'')\ot e_i+\cdots.\cr}
$$
Hence $vt_i=u'v''t_i+v't_iu''$.
{}From this and the hypothesis of the induction,
$$
\eqalignno{
[f_i,\vp S^{-1}(u)]&=[f_i,\vp S^{-1}(u')\vp S^{-1}(u'')]\cr
&=[f_i,\vp S^{-1}(u')]\vp S^{-1}(u'')+\vp S^{-1}(u')[f_i,\vp S^{-1}(u'')]\cr
&={{\vp S^{-1}(v')t_i}\over{\qii-\qi}}\cdot\vp S^{-1}(u'')+
 \vp S^{-1}(u')\cdot{{\vp S^{-1}(v'')t_i}\over{\qii-\qi}}\cr
&={{\vp S^{-1}(u'v''t_i+v't_iu'')}\over{\qii-\qi}}
={{\vp S^{-1}(v)t_i}\over{\qii-\qi}}.&{\hbox{Q.E.D.}}\cr}
$$

\vskip2mm
\nd We return to the proof of Lemma 6.3.
We write $C_{\beta}=\sum_rx^{\beta}_r\ot y^{-\beta}_r$.
We shall show the equality of (6.7) by applying $1\ot (u,\cdot)$ to
the both sides of (6.7), where $u\in U^+_{\beta}$ and
$(\,,\,)$ is the Killing form.
$$
\eqalign{
&1\ot(u,\cdot)[f_i\ot1,\wtil C_{\beta+\al_i}]\cr
&\qq=(\sum_rf_i\cdot\vp S^{-1}((u,y^{-\beta-\al_i}_r)x^{\beta+\al_i}_r)-
\vp S^{-1}((u,y^{-\beta-\al_i}_r)x^{\beta+\al_i}_r)\cdot f_i)\ot1\cr
&\qq=[f_i,\vp S^{-1}(u)].\cr}
$$
On the other hand, by Lemma 6.4 and the properties of the Killing form,
$$
\eqalignno{
\{1\ot(u,\cdot)\}\wtil C_{\beta}(t_i\ot f_i)
&=\sum_r\vp S^{-1}(x^{\beta}_r)t_i\ot(u,y^{-\beta}_rf_i)\cr
&=\sum_r\vp S^{-1}(x^{\beta}_r)t_i\ot(\Del(u),y^{-\beta}_r\ot f_i)\cr
&=\sum_r\vp S^{-1}(x^{\beta}_r)t_i\ot(vt_i,y^{-\beta}_r)(e_i,f_i)\cr
&=\sum_r\vp S^{-1}((vt_i,y^{-\beta}_r)x^{\beta}_r)t_i/(\qii-\qi)\cr
&=\vp S^{-1}(v)t_i/(\qii-\qi).&{\hbox{Q.E.D.}}\cr}
$$

Let us show (6.5).
Multiplying  $q^{(\beta+\al_i,\beta)}(k_{-\beta-\al_i}\ot k_{\beta})$ to
the both sides of (6.7), we obtain
$$
\eqalign{
&q^{(\beta+\al_i,\beta+\al_i)}(f_i\ot t^{-1}_i)
(k_{-\beta-\al_i}\ot k_{\beta+\al_i})\wtil C_{\beta+\al_i}\cr
&=q^{(\beta+\al_i,\beta+\al_i)}
(k_{-\beta-\al_i}\ot k_{\beta+\al_i})\wtil C_{\beta+\al_i}(f_i\ot t^{-1}_i)
+q^{(\beta,\beta)}(k_{-\beta}\ot k_{\beta})\wtil C_{\beta}(1\ot f_i).\cr}
\eqno(6.9)
$$
{}From (6.9) and the formula (B6)
$q^{-H}(f_i\ot t^{-1}_i)=(f_i\ot1)q^{-H}$,
we get
$$
\eqalign{
&(f_i\ot 1)\cdot q^{-H}q^{(\beta+\al_i,\beta+\al_i)}
(k_{-\beta-\al_i}\ot k_{\beta+\al_i})\wtil C_{\beta+\al_i}\cr
&=q^{-H}\{q^{(\beta+\al_i,\beta+\al_i)}
(k_{-\beta-\al_i}\ot k_{\beta+\al_i})\wtil C_{\beta+\al_i}(f_i\ot t^{-1}_i)
+q^{(\beta,\beta)}(k_{-\beta}\ot k_{\beta})\wtil C_{\beta}(1\ot
f_i)\}.\cr}
$$
{}From the presentation (B4), we  have,
$$
(f_i\ot1)\wtil\UR=\wtil\UR(f_i\ot t^{-1}_i+1\ot f_i)=\wtil\UR\Delr(f_i).
$$
{}From this, we get (6.5). \hfill{Q.E.D.}

\vskip3mm
\nd{\sl 6.2. Projector $\Gamma$\q}
We set ${\cal C}=\sum_{\beta\in Q_+}q^{(\beta,\beta)}
(k_{\beta}^{-1}\ot k_{\beta})C_{\beta}\in
\widehat U\widehat\ot\widehat U$ and set
$\wtil{\cal C}=(\vp S^{-1}\ot1){\cal C}$.
{}From the result of [T](Sect.4), we know that
$$
\eqalign{
&{\cal C}^{-1}=\sum_{\beta\in Q_+}q^{(\beta,\beta)}
(1\ot k_{\beta})(S\ot1)(C_{\beta}),\cr
&\wtil{\cal C}^{-1}=(\vp S^{-1}\ot1){\cal C}^{-1}=
\sum_{\beta\in Q_+}q^{(\beta,\beta)}
(1\ot k_{\beta})(\vp\ot1)(C_{\beta}).\cr}
$$
We write $\wtil{\cal C}^{-1}=\sum_ka_k\ot b_k$,
where $a_k\in \bup$ and $b_k\in\ulow$ and set
$$
\Gamma=\sum_kS^{-1}(b_k)a_k\in \widehat B.
$$
This is well-defined as an endomorphism of objects in $\catob$.

\proclaim Proposition 6.4.
For any $\lm\in \ttt^*$, we have
$$
\Gamma^2=\Gamma,\qq
\Gamma\cdot H(\lm)=\FF u_{\lm},\eqno(6.10)
$$
and in particular, $\Gamma u_{\lm}=u_{\lm}$.
\par

\vskip2mm
\nd{\sl Proof.}
{}From (6.9)
we obtain
$(f_i\ot t^{-1}_i)\wtil{\cal C}=\wtil{\cal C}\Delr(f_i)$ for any $i$,
and then
$\wtil{\cal C}^{-1}(f_i\ot t^{-1}_i)=\Delr(f_i)\wtil{\cal C}^{-1}.$
Thus
$$
\sum a_kf_i\ot b_kt^{-1}_i=\sum f_ia_k\ot t^{-1}_ib_k+a_k\ot f_ib_k.\eqno(6.11)
$$
Applying $m\circ\sigma(1\ot S^{-1})$ to the both-side of (6.11),
where $\sigma:a\ot b\mapsto b\ot a$ and $m$ is a multiplication, we have
$$
\eqalign{
\sum t_iS^{-1}(b_k)a_kf_i&
=\sum S^{-1}(b_k)t_if_ia_k-S^{-1}(b_k)t_if_ia_k=0.\cr}
$$
Thus $\Gamma\cdot f_i=0$ for any $i\in I$.
{}From this and Proposition 2.3, we get (6.10).
\hfill{Q.E.D.}

\vskip3mm
\nd{\bf Example.} For $\ge=\ssl_2$, we have
$$
\Gamma=\sum_{n\geq0}q^{{1\over2}n(n-1)}
(-1)^nf^{(n)}{e''}^n.\eqno(6.12)
$$
Note that  an element similar to (6.12) is introduced in [K1].

\par

\par
\vskip5mm
\nd{\bf Appendix A.}

\nd We list several formulae for the operations in Sect.1,
which are analogs of the formula for a
Hopf algebra.
$$
\eqalignno{
&(1\ot m)(1\ot\vp\ot 1)(1\ot \Delb)\Delr(X)=X\ot 1\qq(X\in B)&(A1)\cr
&(m\ot 1)(1\ot\vp\ot 1)(1\ot \Delb)\Delr(X)=1\ot X\qq(X\in B)&(A2)\cr
&(1\ot m)(1\ot\sigma)(1\ot\vpi\ot 1)(\Delb\ot 1)\Dell(X)
=X\ot1\qq(X\in \ovl B)&(A3)\cr
&(m\ot 1)(\sigma\ot1)(1\ot\vpi\ot1)(\Delb\ot 1)\Dell(X)
=1\ot X\qq(X\in \ovl B)&(A4)\cr
&(1\ot m)(1\ot\vp\ot1)(\Dell\ot1)\Delb(X)=X\ot1\qq(X\in U)&(A5)\cr
&(1\ot m)(\vp\ot1\ot1)(1\ot\Delr)\Delb(X)=1\ot X\qq(X\in U)&(A6)\cr
&m(\vp\ot1)\Delb(X)=\varepsilon(X)\qq(X\in U)&(A7)\cr
&(1\ot\varepsilon)\Delr(X)=X\ot1\qq(X\in B)&(A8)\cr
&(\varepsilon\ot1)\Dell(X)=1\ot X\qq(X\in \ovl B)&(A9)\cr
&\Dell\vpi(X)=(a^{-1}\ot\vpi)\sigma\Delr(X)
\qq(X\in B)&(A10)\cr
&\Delr\vp(X)=(\vp\ot a)\sigma\Dell(X)
\qq(X\in \ovl B),&(A11)\cr
&\Delb(X)=(1\ot \vp S^{-1})\Del(X)\qq(X\in U^+),&(A12)\cr
&\Delb(X)=(\vpi S\ot 1)\Del(X)\qq(X\in U^-),&(A13)\cr}
$$
where $\sigma:a\ot b\lar b\ot a$ and $m$ is a multiplication
$m:a\ot b\lar ab$.

These are obtained by direct calculations. We shall show, for example, (A1).
First we show for generators; this is trivial.
Next, we assume that $x$ and $y\in B$ satisfy (A1) and
write  $(1\ot\Delb)\Delr(u)=\sum u_{(1)}\ot u_{(2)}\ot u_{(3)}$.
Then we have
$$
\eqalign{
&(1\ot m)(1\ot\vp\ot 1)(1\ot \Delb)\Delr(xy)
=(1\ot m)\sum x_{(1)}y_{(1)}\ot \vp(y_{(2)})\vp(x_{(2)})\ot x_{(3)}y_{(3)}\cr
&=\sum x_{(1)}y_{(1)}\ot \vp(y_{(2)})\vp(x_{(2)})x_{(3)}y_{(3)}
=\sum xy_{(1)}\ot \vp(y_{(2)})y_{(3)}=xy\ot1.\cr}
$$
Thus we get (A1).
\beginsection Appendix B.

In this appendix, we recall the theory of
the universal R-matrix of $U$ (see [D1],[T]).

Recall that for the Hopf algebra $(U,\Del,S,\vep)$ the universal R-matrix
$\UR$ is an element which enjoys the following properties ([D1],[T]),
$$
\eqalignno{
&\UR\Del(x)=\Del'(x)\UR{\hbox{ for any }}x\in U,&(B1)\cr
&(\Del\ot 1)\UR=\UR_{13}\UR_{23},\q(1\ot\Del)\UR=\UR_{13}\UR_{12},&(B2)\cr
&(\vep\ot{\hbox{id}})\UR=1\ot1=({\hbox{id}}\ot\vep)\UR,\q
(S\ot{\hbox{id}})\UR=\UR^{-1}=({\hbox{id}}\ot S)\UR.&(B3)\cr}
$$
We need some preparation to write down the explicit form of $\UR$.
Let $\widehat U\widehat\ot\widehat U$ be a weight completion of $U\ot U$
as in Sect.1.
Let $H\in \ttt\ot\ttt$ be a canonical element with respect to
the invariant bilinear form on $\ttt$.
We extend the algebra $\widehat U\widehat\ot \widehat U$ by adding
formal elements $q^{\pm H}$ with the following properties;
$$
\eqalignno{
&q^H\cdot q^{-H}=q^{-H}\cdot q^H=1\ot1,\q
q^{\pm H}(q^h\ot q^{h'})=(q^h\ot q^{h'})q^{\pm H},&(B4)\cr
&q^{\pm H}(e_i\ot1)=(e_i\ot t^{\pm}_i)q^{\pm H},\q
 q^{\pm H}(1\ot e_i)=(t^{\pm}_i\ot e_i)q^{\pm H},&(B5)\cr
&q^{\pm H}(f_i\ot1)=(f_i\ot t^{\mp}_i)q^{\pm H},\q
 q^{\pm H}(1\ot f_i)=(t^{\mp}_i\ot f_i)q^{\pm H},&(B6)\cr
&(\Del\ot1)q^{\pm H}=q^{\pm H_{13}}q^{\pm H_{12}},
\q(1\ot \Del)q^{\pm H}=q^{\pm H_{13}}q^{\pm H_{23}},&(B7)\cr}
$$
where $q^{\pm H_{ij}}$'s are  elements
corresponding  to $q^{\pm H}$ on
the $i$-th and the $j$-th components in tensor products
and they commute with each other. Thus, for
example, we identify $q^{H_{12}}$ with $q^H\ot1$.
We denote this algebra by $(\widehat U\widehat\ot \widehat U)^{\widehat{}}$.
{}From the property (B7), we can also extend $\Del\ot1$ and $1\ot\Del$ to the
algebra homomorphism $(\widehat U\widehat\ot \widehat U)^{\widehat{}}
\lar (\widehat U\widehat\ot \widehat U\widehat\ot \widehat U)^{\widehat{}}$.
More generaly, we can extend $\widehat U^{\widehat\ot n}$ to
$(\widehat U^{\widehat\ot n})^{\widehat{}}$ by
adding $q^{\pm H_{ij}}$ ($1\leq i<j\leq n$).

By using the Killing form (see Sect.3) we can
carry out  Drinfeld's quantum double construction formally and get an explicit
presentation of $\UR$,
$$
\UR=q^{-H}\sum_{\beta\in Q_+}q^{(\beta,\beta)}
(k_{\beta}^{-1}\ot k_{\beta})C_{\beta}\in
(\widehat U\widehat\ot\widehat U)^{\widehat{}},\eqno(B8)
$$
where $k_{\beta}$ is an element of $T$
given by $k_{\beta}=\prod_jt^{m_j}_j$ for $\beta=\sum_jm_j\al_j$ and
$C_{\beta}$ is a canonical element of $U^+_{\beta}\ot U^-_{-\beta}$
with respect to the Killing form.

Here, for $U$-modules $V$ and $W$,
$q^{\pm H}$ can be regarded as an element of ${\hbox{End}}(V\ot W)$
given by $q^{\pm H}(u\ot v)=q^{\pm(\xi,\eta)}(u\ot v)$,  ($u\in V_{\xi}$ and
$v\in W_{\eta}$). (See [Kac] Sect.2).
In such consideration,  $\UR$ makes sense as an endomorphism of
tensor products of $U$-modules.
For vectors $u$ and $v$ as above we get,
$$
\eqalign{
&q^{-H+(\beta,\beta)}(k^{-1}_{\beta}\ot k_{\beta})C_{\beta} (u\ot v)\cr
&\qq=q^{-H-(\beta,\beta)}C_{\beta}(k^{-1}_{\beta}\ot k_{\beta}) (u\ot v)
=q^{-H-(\beta,\beta)+(\beta,\eta-\xi)}C_{\beta}(u\ot v)\cr
&\qq
=q^{-(\xi+\beta,\eta-\beta)-(\beta,\beta)+(\beta,\eta-\xi)}C_{\beta}(u\ot v)
=q^{-(\xi,\eta)}C_{\beta}(u\ot v).\cr}
$$
Therefore, we obtain
$$
\UR(u\ot v)=q^{-(\xi,\eta)}\sum_{\beta}C_{\beta}(u\ot v).\eqno(B9)
$$

\vskip3mm
When $\ge$ is an affine Lie algebra, we set
$$
\UR'(z)=q^{-H+c\ot d+d\ot c}\sum_{\beta\in Q_+}q^{(\beta,\beta)}
(z^{\lan d,\beta\ran}k_{\beta}^{-1}\ot k_{\beta})C_{\beta},\eqno(B10)
$$
where $c$ is a canonical central element of
$\ge$ and $d$ is a scaling element of $\ge$.
This is used  to describe the image of the universal R-matrix
onto a tensor product of affinization for
finite dimensional $U'$-modules. (see [FR],[IIJMNT].)
\beginsection Reference

\def\CMP{\sl Commum.Math.Phys.}
\def\IJMP{\sl Int.J.Mod.Phys.}

\item{[A]} {Abe E}, {Hopf algebras},
Cambridge Univ. Press (1980).
\item{[D1]}{Drinfeld V G, Quantum Groups}, {Proceedings of ICM Barkeley},
(1986) 798--820.
\item{[D2]} {Drinfeld V G, On almost co-commutative Hopf algebras},
{\sl Leningrad Math. J.}{\bf 1} {(1990) 321--431}.
\item{[DFJMN]} Davies B, Foda O, Jimbo M, Miwa T and Nakayashiki A,
Diagonalization of the XXZ Hamiltonian by vertex operators,
{\CMP}, {\bf 151}, (1993) 89--153.
\item{[DJO]} {Date E, Jimbo M and Okado M,
Crystal base and $q$-vertex operators},
{\sl Osaka Univ. Math. Sci. preprint}{\bf 1} (1991), to appear in {\CMP}
\item{[FR]} Frenkel I B and Reshetikhin N Yu,
Quantum affine algebras and holonomic difference equations,
{\CMP}, {\bf 149}, (1992) 1--60.
\item{[IIJMNT]} Idzumi M, Iohara K, Jimbo M, Miwa T, Nakashima T and
Tokihiro T, Quantum affine symmetry in vertex models, {\IJMP} A Vol 8,
No.8, (1993) 1479--1511.
\item{[KMN${}^2$] }
 Kang S-J, Kashiwara M, Misra K, Miwa T, Nakashima T and Nakayashiki A,
Affine crystals and vertex models,
{\IJMP},{A7 }Suppl. 1A (1992) 449--484.
\item{[Kac]} Kac V G, Infinite dimensional Lie algebras,
3rd edition, Cambridge Univ. Press (1990).
\item{[KK]} Kac V G and Kazhdan D A, Structure of representations with
highest weight of infinite-dimensional Lie algebras, {\sl Advances in Math.}
{\bf 34} (1979), 97--108.
\item{[K1]} Kashiwara M,
On crystal bases of the $q$-analogue of universal enveloping algebras,
{\it Duke Math. J.},{\bf 63} (1991) 465--516.
 \item{[K2]} Kashiwara M,
Global crystal bases of quantum groups,
{\sl Duke Math. J.}, {\bf 69} (1993) 455--485.
\item{[R]} Rosso, M, Analogues de la forme de Killing et du
th\' eor\` eme d'Harish-Chandra pour les groupes quantiques,
{\sl Ann.scient.\' Ec,Norm. Sup.} {\bf 23} (1990)  445--467.
\item{[S]} Shapovalov N N, On a bilinear form on the universal
enveloping algebra of a complex semisimple Lie algebra, {\sl Funkt. Analis.
Appl.}  6 (1972), 307--312.
\item{[T]} Tanisaki T, Killing forms, Harish-Chandra isomorphisms,
and universal $R$-matrices for quantum algebras,
{\IJMP} {A7} {Suppl. 1B (1992) 941--961}.

\par

\end